\documentclass[12pt,preprint]{aastex}

\usepackage{color}

\slugcomment{draft of \today}

\shorttitle{Re-acceleration at Cluster Shocks}
\shortauthors{Kang \& Ryu}

\def\kms{~{\rm km~s^{-1}}}
\def\cm3{~{\rm cm^{-3}}}
\def\yr{~{\rm yr}}

\def\muG{~{\mu\rm G}}

\def\kpc{~{\rm kpc}}

\begin{document}

\title{RE-ACCELERATION MODEL FOR RADIO RELICS WITH SPECTRAL CURVATURE}

\author{Hyesung Kang$^1$ and Dongsu Ryu$^{2,3,4}$}

\affil{$^1$Department of Earth Sciences, Pusan National University, Pusan 46241, Korea: hskang@pusan.ac.kr\\
$^2$Department of Physics, UNIST, Ulsan 44919, Korea: ryu@sirius.unist.ac.kr\\
$^3$Korea Astronomy and Space Science Institute, Daejeon, 34055, Korea}
\altaffiltext{4}{Author to whom all correspondence should be addressed.}

\begin{abstract}
Most of the observed features of radio {\it gischt} relics 
such as spectral steepening across the relic width and power-law-like integrated spectrum
can be adequately explained by diffusive shock acceleration (DSA) model, 
in which relativistic electrons are (re-)accelerated at shock waves induced in the intracluster medium.
However, the steep spectral curvature in the integrated spectrum above 
$\sim 2$ GHz detected in some radio relics such as the Sausage relic in cluster CIZA J2242.8+5301
may not be interpreted by simple radiative cooling of postshock electrons.
In order to understand such steepening,
we here consider a model in which a spherical shock sweeps through and then 
exits out of a finite-size cloud with fossil relativistic electrons. 
The ensuing integrated radio spectrum is expected to steepen much more than predicted for
aging postshock electrons,
since the re-acceleration stops after the cloud-crossing time.
Using DSA simulations that are intended to reproduce radio observations of 
the Sausage relic,
we show that both the integrated radio spectrum and the surface brightness profile can 
be fitted reasonably well, if a shock of speed, $u_s \sim 2.5-2.8\times 10^3 \kms$, and 
sonic Mach number, $M_s \sim 2.7-3.0$, traverses a fossil cloud for $\sim 45$ Myr and 
the postshock electrons cool further for another $\sim 10$ Myr.
This attempt illustrates that steep curved spectra of some radio gischt relics
could be modeled by adjusting the shape of the fossil electron spectrum
and adopting the specific configuration of the fossil cloud.
\end{abstract}

\keywords{acceleration of particles --- cosmic rays --- galaxies: clusters: general --- shock waves}

\section{INTRODUCTION}

Radio relics are diffuse radio sources that contain relativistic electrons with the Lorentz factor of $\gamma_e\sim 10^4$,
radiating synchrotron emission in the magnetic fields of order of $\mu$G in galaxy clusters
\citep[see, e.g.,][for reviews]{feretti12,brunetti2014}.
After the classification scheme of \citet{kempner04}, they are often divided 
into two main groups  according to their
origins and observed properties: {\it radio gischt} and {\it AGN relic/radio phoenix}.
Radio gischt relics are thought to be produced by merger-driven shocks.
They are characterized by elongated shape, radio spectrum steepening behind the hypothesized shock,
power-law-like integrated radio spectrum, and high polarization fraction \citep[e.g.][]{ensslin98,brug12}.
They are found mainly in the periphery of merging clusters.
Giant radio relics such as the Sausage relic in CIZA J2242.8+5301 and the Toothbrush relic 
in 1RXS J0603.3 are typical examples of radio gischt \citep{vanweeren10, vanweeren2012}.
On the other hand, AGN relics are radio-emitting relativistic plasmas ejected from radio-loud AGNs,
and they turn into {\it radio ghosts} (undetectable in radio) quickly due to fast electron cooling when their
source AGNs are extinct \citep{ensslin99}.
Radio ghosts can be reborn later as radio phoenixes, if cooled electron plasmas are compressed and re-energized by
structure-formation shocks \citep{ensslin01}.
Radio phoenixes have roundish {or ring-like} shape and steep-curved integrated radio spectrum of aged electron populations,
and they are found near their source galaxies in the cluster center region \citep[e.g.][]{slee01,vanweeren11b}.
The relics in A2443 and A1033 are radio phoenixes \citep{clarke13,deGasperin15}.

In the diffusive shock acceleration (DSA) model for radio {gischt} relics, 
relativistic electrons are accelerated or re-accelerated at cluster shocks that are
driven by supersonic motions associated with mergers of sub-structures or infall of the warm-hot
gas along filaments into the hot intracluster medium (ICM) \citep[e.g.][]{ensslin98,hoeft08,vazza12, hong14}.
Although the injection of seed electrons into Fermi first
order process at {\it weak} cluster shocks has not been fully understood \citep[e.g.][]{kang14, guo14},
it is now expected that the {\it in situ} injection/acceleration from the ICM thermal plasma and/or 
the re-acceleration of fossil relativistic electrons
may explain the radio flux level of observed radio gischt relics \citep[e.g.][]{kang12,vazza2015}.
In particular, the presence of AGN relics and radio phoenixes implies that the ICM may host clouds of 
aged relativistic electrons with $\gamma_c\la 300$. 
Re-acceleration of those electrons by cluster shocks as the origin of radio relics 
has been explored by several authors \citep[e.g.][]{kangryu11, kang12, pinzke13}.

In the re-acceleration model for radio gischt relic, it is assumed that the shock propagates in the ICM 
thermal plasma that contains an additional population of suprathermal electrons 
with dynamically insignificant nonthermal pressure 
{\citep[e.g.][]{ensslin98, kang12,pinzke13}.}
The role of those fossil electrons is to provide seed electrons to the DSA process. 
In the compression model for radio phoenix, on the other hand,
the fossil radio plasma does not mix with the background gas and has high buoyancy and high sound speed
\citep{ensslin01}.
So the radio plasma is compressed only adiabatically, when a shock wave sweeps through the radio bubble.
{The shock passage through such hot radio plasma is expected to result in
a filamentary or toroidal structure \citep{ensslin02,pfrommer11},}
which is consistent with the observed morphology of radio phoenixes \citep{slee01}.

According to cosmological hydrodynamical simulations for large-scale
structure formation, shocks are ubiquitous in the ICM with the mean separation of $\sim 1$ Mpc 
between shock surfaces and with the mean life time of $t_{\rm dyn}\sim 1$ Gyr 
\citep[e.g.,][]{ryu03,pfrommer2006,skillman2008,vazza09}.
So it is natural to expect that actively merging clusters would contain at least several shocks and 
associated radio relics.
However, the fraction of X-ray luminous merging clusters 
hosting radio relics is observed to be order of $\sim 10$ \% \citep{feretti12}.
In order to reconcile such rarity of observed radio gischt relics with the frequency of shocks 
estimated by those structure formation simulations,
\citet{kangryu15} (Paper I) proposed a scenario in which shocks in the ICM may light up as radio relics only when they 
encounter clouds of fossil relativistic electrons left over from either
radio jets from AGNs or previous episodes of shock/turbulence acceleration.

In the basic DSA model of steady planar shock with constant postshock magnetic field, 
the electron distribution function at the shock location becomes a power-law
of $f_e(p,r_s)\propto p^{-q}$ with slope $q=4M_s^2/(M_s^2-1)$,
while the volume-integrated electron spectrum behind the shock becomes $F_e(p)\propto p^{-(q+1)}$ \citep{dru83}.
Then, the synchrotron spectrum at the shock becomes a power-law of
$j_{\nu}(r_s)\propto \nu^{-\alpha_{\rm sh}}$ with the {\it shock index}, $\alpha_{\rm sh} = (M_s^2+3)/2(M_s^2-1)$,
while the volume-integrated radio spectrum becomes $J_{\nu} \propto \nu^{-A_{\nu}}$ 
with the {\it integrated index}, $A_\nu=\alpha_{\rm sh}+0.5$, above the break frequency $\nu_{\rm br}$
\citep[e.g.][]{ensslin98,kang11}. Here, $M_s$ is the sonic Mach number of the shock.

The simple picture of DSA needs to be modified in real situations. 
In the re-acceleration model, for instance, 
if the fossil electrons dominate over the electrons injected from ICM plasma, 
the ensuing electron spectrum must depend on the shape of the fossil electron spectrum 
and may not be a single power-law.
In addition, if the shock acceleration duration is less than $\sim 100$ Myr, 
the integrated radio spectrum cannot be a simple power-law, 
but instead it steepens gradually over the frequency range of $0.1-10$ GHz, 
since the synchrotron break frequency falls to $\nu_{\rm br}\sim 1$ GHz or so \citep{kang15a}.
\citet{kang15b} demonstrated that, even with a pure {\it in situ} injection model, 
both the electron spectrum and the ensuing radio spectrum could depart from simple power-law forms
in the case of spherically expanding shocks with varying speeds and/or nonuniform magnetic field profiles.
In Paper I, we showed that the re-acceleration of fossil electrons at spherical shocks expanding through 
cluster outskirts could result in the curved integrated spectrum.

In some radio relics, the integrated radio spectra exhibit the steepening above $\sim 2$ GHz,
but much stronger than predicted from simple radiative cooling of shock-accelerated
electrons in the postshock region.
For instance, \citet{trasatti15} suggested that 
the integrated spectrum of the relic in A2256 could be fitted by a broken power-law with
$A_\nu \approx 0.85$ between 0.35 GHz and 1.37 GHz and $A_\nu \approx 1.0$ between 1.37 GHz and 10.45 GHz.
Recently, \citet{stroe16} showed that 
the integrated spectral index of the Sausage relic increases from $A_\nu \approx 0.9$ below $2.5$ GHz
to $A_\nu \approx 1.8$ above $2.0$ GHz,
while that of the Toothbrush relic increases from $A_\nu \approx 1.0$ below $2.5$ GHz
to $A_\nu \approx 1.4$ above $2.0$ GHz.
Note that $A_\nu \approx 0.9-1.0$ for the low frequency spectrum of the Sausage and Toothbrush relics 
is larger than the inferred shock index, $\alpha_{\rm sh}\approx 0.6-0.7$,
while $A_\nu \approx 1.4-1.8$ for the high frequency part is also larger than $\alpha_{\rm sh}+0.5 \approx 1.1-1.2$ \citep{vanweeren2012,stroe14}.
In particular, in the case of the Sausage relic, the steepening of $J_{\nu}$ is much stronger 
than expected for aging postshock electrons.
This demonstrates that the simple relation of $A_\nu=\alpha_{\rm sh}+0.5$ should be applied only with caution
in interpreting observed radio spectra.
The picture of DSA just based on shock compression and radiative cooling should be too simple to be applied to real situations, which could be complicated 
by additional elements such as the presence of pre-exiting electron population and 
the variations of shock dynamics and magnetic field amplification.
{It was pointed that the Sunyaev-Zeldovich (SZ) effect can induce a steepening at high frequencies.
\citet{basu15} argued that the effect may reduce the radio flux by a factor of two or so
at $\nu \sim 30$ GHz for the case of of the Sausage relic.
On the other hand, the observations require a reduction of a factor of several.
So although the detailed modeling still has to be worked out,
the SZ effect alone would not be enough to explain the observed steepening.}

In an attempt to reproduce the observed spectrum of the Sausage relic,
Paper I showed that the integrated spectrum estimated from the DSA simulations of spherical shocks
expanding in the cluster outskirts 
with the `acceleration' age, $t_{\rm age}\la 60-80$~Myr,
steepens only {\it gradually} over $0.1-10$ GHz.
But the abrupt increase of $A_{\nu}$ above $\sim 2$ GHz detected in the Sausage relic could not be explained,
implying that additional physical processes would operate for electrons with $\gamma_e \ga 10^4$.
In this study, we propose a simple but natural re-acceleration scenario, in which 
the shock passes through a finite-size cloud of fossil electrons and runs ahead of the postshock volume
of radio-emitting electrons.
Since the supply of seed electrons is stopped outside the cloud,
the shock no longer efficiently accelerate electrons.
Then, the integrated radio spectrum of the relic steepens beyond radiative cooling alone,
and the shock front and the radio relic do not coincide spatially with each other.

Another observational feature that supports the re-acceleration scenario is 
nearly the uniform surface brightness along thin elongated structures observed in some relics, such as
the Sausage relic and the Toothbrush relic \citep{vanweeren10, vanweeren2012}.
Using numerical simulations of merging cluster, \citet{vanweeren11a} demonstrated that a merger-driven bow
shock would generate the surface brightness profile that peaks in the middle and decreases 
along the length of the relic away from the middle, 
if the electrons are injected/accelerated at the shock and occupy a portion of spherical shell.
So their study implies that the pure {\it in situ} injection picture may not explain 
the uniform surface brightness profile.
On the other hand, \citet{vanweeren10} and \citet{kang12} showed that
the radio flux density profile with the observed width of $\sim 55$ kpc for the Sausage relic
can be fitted by a patch of cylindrical shell with radius $\sim 1.5$ Mpc, 
which is defined by a length of $\sim 2$ Mpc and an `extension angle' of $\psi\approx 10^{\circ}$.
Although rather arbitrary and peculiar, such a configuration could yield uniform radio flux density 
along the length of the radio relic.
In Paper I, we proposed that such a geometrical structure can be produced, if a spherical shock
propagates into a long cylindrical volume of fossil electrons and the re-accelerated population of fossil electrons
dominates over the injected electron population \citep[see also Figure 1 of][]{kang15b}.
Here, we adopt the same geometrical structure for the radio-emitting volume
in the calculation of the surface brightness profile,
but assuming that the shock has existed and runs ahead of the volume.

In the next section, the numerical simulations and the shock models, designed to reproduced the Sausage relic, are described.
In Section 3, our results are compared with the observations of the Sausage relic.
A brief summary is followed in Section 4.

\section{NUMERICAL CALCULATIONS}

The numerical setup for DSA simulations, the basic features of DSA and synchrotron/inverse-Compton (iC) cooling,
and the properties of shocks and magnetic fields in the ICM were explained in details in Paper I.
Here only brief descriptions are given. 
 
\subsection{DSA Simulations for 1D Spherical Shocks}

The diffusion-convection equation for the relativistic electron population is solved 
in the one-dimensional (1D) spherical geometry:
\begin{eqnarray}
{\partial g_e\over \partial t}  + u {\partial g_e \over \partial r}
= {1\over{3r^2}} {{\partial (r^2 u) }\over \partial r} \left( {\partial g_e\over
\partial y} -4g_e \right)  
+ {1 \over r^2}{\partial \over \partial r} \left[r^2 D(r,p)  
{\partial g_e \over \partial r} \right]
+ p {\partial \over {\partial y}} \left( {b\over p^2} g_e \right),
\label{diffcon}
\end{eqnarray}
where $g_e(r,p,t)=f_e(r,p,t) p^4$ is the pitch-angle-averaged phase space distribution function
of electrons and $y \equiv \ln(p/m_e c)$ with the electron mass $m_e$ and
the speed of light $c$ \citep{skill75}.
The background flow velocity, $u(r,t)$, is obtained by solving the usual gasdynamic conservation equations 
in the test-particle limit where the nonthermal pressure is assumed to be negligible.

The spatial diffusion coefficient for relativistic electrons is assumed to have 
the following Bohm-like form,
\begin{equation}
D(r,p) = 1.7\times 10^{19} {\rm cm^2s^{-1}} \left({ B(r)\over 1\muG}\right)^{-1} 
\left({p \over m_e c}\right).
\label{Bohm}
\end{equation}
Then, the cutoff Lorentz factor in the shock-accelerated electron spectrum is given by
\begin{equation}
\gamma_{e, {\rm eq}} \approx 10^9 \left({u_s \over {3000 \kms}}\right) \left(B_1 \over 1\muG \right)^{1/2} \left((1 \muG)^2 \over {B_{\rm e,1}^2 + B_{\rm e,2}^2}\right)^{1/2},
\label{gammaeq}
\end{equation}
where $B_{\rm e}^2 \equiv B^2 + B_{\rm rad}^2$ is the `effective' magnetic field strength
which accounts for both synchrotron and iC losses \citep{kang15a}.
Hereafter, the subscripts ``1'' and ``2'' are used to indicate the preshock and postshock quantities, respectively.
Note that for electrons with $\gamma_{e, {\rm eq}} \sim 10^8$, the diffusion length is 
{$D(\gamma_e)/u_s\sim 2~{\rm pc}$ and
the diffusion time is $D(\gamma_e)/u_s^2\sim 600\yr$, if $B\sim 1\muG$ and $u_s\sim 3\times10^3 \kms$.}
So in effect electrons are accelerated almost {\it instantaneously} to the cutoff energy 
at the shock front.

The radiative cooling coefficient, $b(p)$, is calculated from the cooling time scale as
\begin{equation}
 t_{\rm rad} (\gamma_e)= {p \over b(p)} =
9.8\times 10^{7} \yr \left({B_{\rm e} \over {5 \muG}}\right)^{-2} 
\left({\gamma_e \over 10^4 }\right)^{-1}.
\label{trad}
\end{equation}
The cooling time scale for the radio-emitting electrons with $\gamma_e=10^3-10^4$ is about 
$10^8-10^9$ yr for the magnetic field strength of a few $\mu$G.

\subsection{Shock Parameters}

For the initial setup, we adopt a Sedov blast wave propagating into a uniform static medium, which can be specified by two parameters, typically,
the explosion energy and the background density \citep[e.g.,][]{ryu91}.
Here, we instead choose the initial shock radius and speed as
$r_{s,i}=1.3~{\rm Mpc}$ and $u_{s,i}=M_{s,i} \cdot c_{s,1}$ (see Table 1).
As in Paper I, the shock parameters are chosen to emulate the shock associated with the Sausage relic;
the preshock temperature is set to be $kT_1 = 3.35$ keV, corresponding to the preshock sound speed of
$c_{s,1}=923\kms$ \citep{ogrean14}.
The background gas is assumed to be isothermal, 
since the shock typically travels only $\sim 200$ kpc,
which is sufficiently small compared to the size of galaxy clusters,
for the duration of our simulations $\la 70$ Myr.

The density of the background gas in cluster outskirts is assumed to decrease as 
$\rho_{\rm up}=\rho_0(r/r_{s,i})^{-2}$.
This corresponds to the so-called beta model for isothamal ICMs,
$\rho(r)\propto [ 1+ (r/r_c)^2 ]^{-3\beta/2}$ with $\beta \sim 2/3$,
which is consistent with typical X-ray brightness profiles of observed clusters \citep{sarazin86}.
Since we neglect the {\it in situ} injection at the shock front (see Section 2.4) and
we do not concern about the absolute radio flux level in this study (see Section 3), 
$\rho_0$ needs not to be specified.

\subsection{Models for Magnetic Fields}

Although the synchrotron cooling and emission of relativistic electrons in radio relics
are determined mainly by postshock magnetic fields, 
little has yet been constrained by observations.
Thus, we consider a rather simple model for postshock magnetic fields as in Paper I.
(1) The magnetic field strength across the shock transition is assumed to increase 
due to the compression of two perpendicular components,
\begin{equation}
B_2(t)=B_1 \sqrt{1/3+2\sigma(t)^2/3}, 
\label{b2}
\end{equation}
where $B_1$ and $B_2$ are the preshock and postshock magnetic field strengths,
respectively, and $\sigma(t)=\rho_2/\rho_1$ is the density compression ratio across the shock.
(2) For the downstream region ($r<r_s$), the magnetic field strength is assumed to scale with the 
gas pressure as
\begin{equation}
B_{\rm dn}(r,t)= B_2(t) \cdot [P_g(r,t)/P_{g,2}(t)]^{1/2},
\label{bd}
\end{equation}
where $P_{g,2}(t)$ is the gas pressure immediately behind the shock.
In effect, the ratio of the magnetic to thermal pressure, that is, the plasma beta, is assumed to be constant downstream of the shock.

\subsection{Fossil Electron Cloud}

In Paper I, we explored a scenario in which a shock in the ICM lights up as a radio relic
when it encounters a cloud that contains fossil relativistic electrons, 
as described in the Introduction.
In this study, we consider a slightly modified scenario in which
a spherical shock passes across a fossil electron cloud with 
width $L_{\rm cloud}\sim 100$ kpc.
Then, the shock separates from and runs ahead of the downstream radio-emitting electrons 
after the crossing time, 
$t_{\rm cross}\sim L_{\rm cloud}/u_s\approx 32.6~{\rm Myr} \cdot (L_{\rm cloud}/100\kpc)(u_s/3000\kms)^{-1} $.
We assume that the downstream volume with relativistic electrons has the same geometric structure as illustrated in
Figure 1 of \citet{kang15b}, except that the shock is detached from and moves ahead the radio-emitting volume.
In other words, the re-acceleration of seed electrons operates only during $t_{\rm cross}$, that is, between the time of entry into and the time of exit 
out of the fossil electron cloud, and then, the downstream electrons merely cool radiatively up to $t_{\rm age}> t_{\rm cross}$, leading
to the steepening of the integrated radio spectrum.
Hereafter, the `age', $t_{\rm age}$, is defined as the time since the shock enters into the cloud.

The fossil electrons are assumed to have a power-law spectrum with exponential cutoff,
\begin{equation}
f_{\rm fossil}(p)=f_0\cdot \left(p \over p_{\rm inj}\right)^{-s} \exp \left[ - \left({\gamma_e \over \gamma_{e,c}} \right)^2 \right],
\label{fexp}
\end{equation}
with $s=4.0-4.2$ and $\gamma_{e,c}=10^3-10^4$ for the slope and the cutoff Lorentz factor, respectively.
Again the normalization factor, $f_0\cdot p_{\rm inj}^s$, is arbitrary in our calculations.
This could represent fossil electrons that have cooled down to $\sim \gamma_{e,c}$ from much higher energies
for $(0.1-1) \times t_{\rm dyn}\approx 0.1 - 1$ Gyr.
As described in Paper I (also in the Introduction),
one can think of several possible origins for such fossil electrons in the ICM:
(1) old remnants (radio ghosts) of radio jets from AGNs, (2) electron populations that were accelerated by 
previous shocks and have cooled down $\gamma_{e,c}$, and
(3) electron populations that were accelerated by turbulence during merger activities.

In order to focus on the consequences of the fossil electrons,
the {\it in situ} injection is suppressed; that is, we assume that the {\it in situ} injected and accelerated population is negligible compared to the re-accelerated population of the fossil electrons.
We also assume that the nonthermal pressure of the fossil electrons is dynamically insignificant,
thus, the sole purpose of adopting the fossil electrons is to supply seed electrons into the DSA process
at the shock.
Finally, we assume that the background gas has $\gamma_g=5/3$.

\section{RESULTS OF DSA SIMULATIONS}

\subsection{Model Parameters}

We consider several models whose parameters are summarized in Table 1.
In all the models, the preshock temperature is fixed at $kT=3.35$ keV,
and the preshock magnetic field strength at $B_1=2.5\muG$ with the immediate postshock 
magnetic field strength, $B_2(t)\approx 6.0-6.3\muG$ during 60 Myr.
For the fiducial model, M3.0C1, 
the initial shock speed is $u_{s,i}=2.8\times 10^3\kms$, corresponding to the initial shock Mach number $M_{s,i}=3.0$, at the onset of simulation,
the width of the fossil electron cloud is $L_{\rm cloud}=131$ kpc,
and the fossil electron population is specified with the power-law slope $s=4.2$
and the cutoff Lorentz factor $\gamma_{e,c}=10^4$.
The shock speed and Mach number decrease by $\sim 10$ \% or so during 60 Myr in this model as well as in other models in the table (see Paper I).

In the M3.0C1g and M3.0C1s models, different populations of
fossil electrons are considered with $\gamma_{e,c}=10^3$ and $s=4.0$, respectively.
The Mach number dependence is explored with three additional models, M2.5C1, M3.3C1,
and M4.5C1.
The effects of the cloud size are examined in the M3.0C2, M3.0C3, and M3.0C4 models with $L_{\rm cloud}=105$, 155, and 263 kpc, respectively.
In the M3.0C4 model, the shock stays inside the fossil electron cloud until 92 Myr,
so the spectral steepening comes from radiative cooling only.

Lastly, the SC1pex1 model is the same one considered in Paper I, 
in which the initial shock Mach number, $M_{s,i}=2.4$, was
chosen to match the steep spectral curvature above 1.5 GHz in the observed spectrum of the
Sausage relic \citep{stroe14}.
Note that in this model the shock Mach number decreases to $M_s\approx2.1$ at 60 My,
which is lower than $M_{\rm X}\approx 2.54-3.15$ derived from X-ray observations
\citep{akamatsu13,ogrean14}

Figure 1 compares the evolution of the synchrotron emissivity, $j_{\nu}(r,\nu)$, in the M3.0C1,
M3.0C1g, and M3.0C3 models (from top to bottom panels) at age $t_{\rm age}=$ 18, 45, 55,
and 66 Myr (from left to right panels).
In the left-most panels at 18 Myr, the shock at $r=r_s$ has penetrated 52 kpc into the cloud in the three models.
The shock is about to exit out of the cloud at 45 Myr in the M3.0C1 and M3.0C1g models
and at 55 Myr in the M3.0C3 model.
In the right-most panels, the edge of the radio emitting region is located behind the shock front ($r/r_s=1$).
The comparison of M3.0C1 and M3.0C1g models indicates 
that the postshock radio emission does not sensitively depend on the cutoff in the
fossil electron spectrum for the range, $\gamma_{e,c}\sim 10^3-10^4$, considered here.
The distributions of $j_{\nu}(r,\nu)$ of the M3.0C1 and M3.0C1g models at 55 Myr look
similar to that of M3.0C3 at 66 Myr except that the downstream volume is slightly larger in the M3.0C3 model.
Note that the fossil electrons in the cloud start to cool from the onset of the simulations.

\subsection{Electron and Radio Spectra}

Figure 2 compares the evolutions of the electron spectrum and the radio spectrum 
in the M3.0C1 (upper four panels) and M3.0C1g (lower four panels) models at the same $t_{\rm age}$
as in Figure 1.
The electron spectrum at the shock, $g_e(r_s,p)$, 
and the volume-integrated electron spectrum, $G_e(p) = p^4 F_e(p) = p^4 \int 4\pi r^2 f_e(r,p) dr$, 
the particle slopes, $q=-d \ln f_e(r_s,p)/d \ln p$ and $Q=-d \ln F_e(p)/d \ln p$,
are shown in the left panels.
The local synchrotron spectrum at the shock, $j_{\nu}(r_s)$,
and the integrated spectrum, $J_{\nu}=\int j_{\nu}(r) dV $, 
the synchrotron spectral indices, $\alpha_{\rm sh}= -d \ln j_{\nu}(r_s)/d \ln \nu $
and $A_{\nu}= -d \ln J_{\nu}/d \ln \nu $, are shown in the right panels.
Here, the plotted quantities, $g_e$, $G_e$, $j_{\nu}$, and $J_{\nu}$, are in arbitrary units.
The shock quantities, $g_e$, $q$, $j_{\nu}$, and $\alpha_{\rm sh}$, are not shown at $t_{\rm age}=$ 55
and 66 Myr, 
since the shock has exited out of the fossil electron cloud.
Note that here the {\it in-situ} injection/acceleration was suppressed in order to focus on
the re-acceleration of fossil electrons (see Section 2.4).

The shape of $G_e(p)$ is significantly different in the two models
in the momentum range of $\gamma_e\sim 10^3 - 10^4$
due to different exponential cutoffs in the initial seed populations.
Once the re-acceleration of seed electrons has stopped at $t_{\rm cross} \sim$ 45 Myr, 
the gradual steepening of the integrated electron spectrum progresses, as can be seen in 
the magenta dashed and dotted lines for both $G_e$ and its slope $Q$.

The behavior of $\alpha_{\rm sh}$ is related with the electron slope as $\alpha_{\rm sh}\approx (q-3)/2$,
which is exact only in the case of 
a single power-law electron spectrum.
Before the shock exits the cloud, the integrated spectral index increases from 
$A_{\nu}\approx (s-3)/2 \approx 0.6$ to $A_{\nu}\approx \alpha_{\rm sh} + 0.5\approx 1.3$
over a broad range of frequency, $\sim (0.01-30)$ GHz (see the magenta solid and long-dashed lines).
But after $t_{\rm cross}$, the integrated index $A_{\nu}$ becomes much steeper at high frequencies
(see the magenta dashed and dotted lines)
due to the combined effects of radiative cooling and the lack of
re-acceleration at the shock front.
{This suggests that the shock breaking out of a finite-size cloud of fossil electrons may explain
the steepening of the integrated spectrum well beyond the radiative cooling alone.}
The comparison of $J_{\nu}$ (or $A_{\nu}$) of the two models shown in the figure indicates 
there is only a marginal difference at low frequencies below 0.5 GHz,
although they have different exponential cutoffs in the seed populations.

\subsection{Surface Brightness Profile}

As in Paper I, the radio intensity or surface brightness, 
$I_{\nu}$, is calculated by adopting the geometric volume of radio-emitting electrons
in Figure 1 of \citet{kang15b}:
\begin{equation}
I_{\nu}(R)= 2 \int_0^h j_{\nu}(r) d {\it h}, 
\label{SB}
\end{equation}
where $R$ is the distance behind the projected shock edge in the plane of the sky,
$r$ is the radial distance from the cluster center,
and $h$ is the path length along line of sights.
Here, we set the extension angle $\psi=12^{\circ}$, 
slightly larger than $\psi=10^{\circ}$ adopted in Paper I,
in order to reproduce the observed width of the Sausage relic.
Note that the radio flux density, $S_{\nu}$, can be obtained by convolving $I_{\nu}$ with a telescope beam as
$S_{\nu}(R) \approx I_{\nu}(R) \pi \theta_1 \theta_2 (1+z)^{-3}$,
if the brightness distribution is broad compared to the beam size of $\theta_1 \theta_2$.

In Figures 3 and 4, the spatial profiles of $I_{\nu}(R)$ at the radio frequency of 0.6 GHz
and the radio spectral index, $\alpha_{0.6}^{1.4}$, 
estimated between 0.6 GHz and 1.4 GHz,
are shown for all the models listed in Table 1.
The radio intensity
$I_{\nu}(R)$ is plotted at three different $t_{\rm age}$ to illustrate the time evolution,
but $\alpha_{0.6}^{1.4}$ is plotted only at the middle $t_{\rm age}$ for the clarity of the figure.
The profile of $I_{\nu}(R)$ in the SC1exp1 model at 62 Myr is 
shown in the green dotted line for comparison.

Figure 3 compares the models with different cloud sizes, $L_{\rm cloud}=105-263$ kpc,
and the models with different fossil electron populations.
Note that $t_{\rm cross}=$ 37, 45, 54, and 92 Myr for M3.0C2, M3.0C1, M3.0C3, and M3.0C4 models, respectively.
In the M3.0C4 model, the shock still remains inside the cloud at the last epoch of $t_{\rm age}=$ 66 Myr.
So the edge of the relic coincides with the projected shock position ($R=0$) and 
the FWHM of $I_{\nu}(R)$ at 0.6~GHz, $\Delta l_{\rm SB}$, 
continues to increase with time in this model.
In the other models, the shock breaks out of the cloud around $t_{\rm cross}$ and
the relic edge lags behind the shock. 
As a result, $\Delta l_{\rm SB}$ does not increase with time after $t_{\rm cross}$. 
The value of $\Delta l_{\rm SB}$ ranges $43-44$~kpc in M3.0C2, $45-50$~kpc in M3.0C1, $46-57$~kpc in M3.0C3,
and $46-57$ kpc in M3.0C4.
In the comparison model SC1exp1, $\Delta l_{\rm SB}\approx51$~kpc.
In most of the models with $L_{\rm cloud}\ge 131$ kpc, $\Delta l_{\rm SB}$ is in a rough agreement with
the observed width of the Sausage relic at 0.6 GHz, reported in \citet{vanweeren10}.

The distance between the shock front and the relic edge after $t_{\rm cross}$ is
\begin{equation}
 l_{\rm shift}\approx 
10~{\rm kpc}\left({u_2 \over {10^3\kms}}\right)\left({{t_{\rm age}-t_{\rm cross}}\over {10~{\rm Myr}}}\right),
\end{equation}
where $u_2 = c_{s,1} M_{s,i}/\sigma$ is the postshock advection speed.
In the models with $L_{\rm cloud}=131$ kpc, $l_{\rm shift}\approx 10-20$ kpc for $t_{\rm age}=55-66$ Myr. 
So it may be difficult to detect such spatial shift between the shock location in X-ray observations
and the relic edge in radio observations with currently available facilities.
{The shift is, for instance, much smaller than the misalignment of $\sim 200$ kpc
between the X-ray shock location and the radio relic edge detected in the Toothbrush relic
\citep{ogrean13}.
We note that \citet{vanweeren2016} pointed that the slope of the X-ray brightness profile changes
at the expected location of the shock associated with the Toothbrush relic.
This should imply that the evidence for a spatial offset between the shock and
this relic is not compelling anyway.}

At $t_{\rm age}=55$ Myr, the shock of initially $M_{s,i} = 3.0$ weakens to $M_s\approx 2.7$,
so the DSA radio index is expected to be $\alpha_{\rm shock}\approx 0.8$ at the shock location.
In Figure 3, we see that $\alpha_{0.6}^{1.4}$ at $t_{\rm age}=55$ Myr (magenta lines) has
$\sim0.75-0.95$ at the edge of the relic and $\sim 1.5-1.8$ at about 60 kpc downstream of the relic edge,
increasing behind the shock due to the electron cooling.
Especially in the models with $L_{\rm cloud}= 103$ and 131 kpc, $\alpha_{0.6}^{1.4}$ at the relic edge
is larger than $\alpha_{\rm shock}$, since the shock is ahead of the relic edge.
This suggests the possibility that the shock Mach number estimated by X-ray observations 
could be slightly larger than that estimated from the radio spectral index,
which is opposite to the tendency of observational data.
In the case of the Toothbrush relic, for instance,
a `radio Mach number' was estimated to be $M_{\rm radio}\approx 3.3-4.6$ from 
$\alpha_{\rm sh}\approx 0.6-0.7$ \citep{vanweeren2012},
while an `X-ray Mach number', $M_{\rm X} \la 2$, was derived from the density/temperature jump \citep{ogrean13}.
The discrepancy in this relic might be understood by the projection effect of multiple shock surfaces, as suggested in \citet{hong15}.

Figure 4 compares the models with different $M_{s,i}$ at three different $t_{\rm age}$ specified in each panel.
The value of $\Delta l_{\rm SB}$ ranges $44-53$ kpc in M2.5C1, $45-50$ kpc in M3.0C1, $40-47$ kpc in M3.3C1,
and $\sim 44$ kpc in M4.5C1.
Since the preshock sound speed and the magnetic field strength are the same in all the models considered here,
the postshock advection speed, $u_2 = c_{s,1} M_{s,i}/\sigma$,
and the postshock magnetic field strength, $B_2=B_1 \sqrt{1/3+2\sigma^2/3}$, are dependent on $M_{s,i}$ and $\sigma$.
The widths are slightly smaller in the M3.3C1 and M4.5C1 models with higher Mach numbers.

\subsection{Volume-Integrated Spectra}

As discussed in Paper I (see the Introduction), the volume-integrated synchrotron spectrum, $J_{\nu}$, 
is expected to have a power-law form, only for a steady planar shock in uniform background medium, 
and only if $t_{\rm age}$ is much longer than $\sim 100$ Myr.
Otherwise, the spectrum steepens gradually around the break frequency of $J_{\nu}$ given by
{
\begin{equation}
\nu_{\rm br}\approx 0.63 {\rm GHz} \left( {t_{\rm age} \over {100 \rm Myr}} \right)^{-2}
 \left( {(5\muG)^2} \over {B_2^2+B_{\rm rad}^2}  \right)^{2} \left( {B_{2} \over {5\muG}} \right).
\label{fbr}
\end{equation}
}
The shock younger than $100$ Myr has $\nu_{\rm br} \ga 0.6$ GHz, 
with the spectral curvature changing over $\sim (0.1-10) \nu_{\rm br}$,
the typical frequency range of radio observations (Paper I).
Thus, the integrated radio spectra of observed radio relics are likely to be curved instead of
being single power-law.

Figures 5 and 6 show $J_{\nu}$ 
for the same models at the same $t_{\rm age}$ as in Figures 3 and 4, respectively.
Again, $J_{\nu}$ is in arbitrary units.
The filled magenta circles represent the data points for the integrated flux of the Sausage relic,
which were taken from Table 3 of \citet{stroe16} and re-scaled to fit by eye the spectra
of the long-dashed lines, except in the M3.0C3 and M3.0C4 models.
For the M3.0C3 model with $L_{\rm cl}=155$ kpc, 
the data points were fitted to the spectrum of the dashed line at 66 Myr.
On the other hand, in the M3.0C4 model with $L_{\rm cl}=263$ kpc, 
the shock is still inside the fossil electron cloud even at 66 Myr,
so the curvature of $J_{\nu}$ at high frequencies, which is induced only by the radiative losses,
is not steep enough to fit the observed flux above 16 GHz.

Among the models shown in Figure 5,
the fiducial model M3.0C1 as well as the M3.0C1s model with slightly flatter fossil electrons
best reproduce the overall spectrum as well as
the steep curvature at high frequencies.
Other models do not reproduce the observed spectrum as good as the two models.
In the comparison model SC1pex1 of Paper I (green dotted line), 
the shock stays inside the fossil electron cloud 
and so the spectral curvature is due to only radiative cooling.
Although the initial Mach number, $M_{s,i}=2.4$, was chosen for this model to explain the steep
spectrum above 1.5~GHz, yet the abrupt increase of the curvature near 1~GHz
could not be reproduced.

Figure 6 shows that the observed spectrum of the Sausage relic can be fitted reasonable well by
all four models with $L_{\rm cl}=131$ kpc and $M_{s,i}=2.5-4.5$
at about 10 Myr after the shock has exited out of the fossil cloud (long-dashed lines).
They generate much better fit compared to the SC1exp1 model.
In M3.3C1 model, for instance, the shock Mach number decrease to $M_s=3.0$ at 
$t_{\rm age}\approx 53$ Myr and the integrated spectrum at that epoch is in a very good agreement
with the observed spectrum.
But considering that the width of $I_{\nu}(R)$ of the M3.0C1 model agrees a bit better
with the observations (see Figure 4),
we designate M3.0C1 as the fiducial model here.

These simulation results demonstrate that the observed spectrum of the Sausage relic with steep curvature
could be explained naturally by the shock passage over a finite-size cloud of fossil electrons
without invoking any additional physical processes other than the synchrotron/iC coolings.

\section{SUMMARY}

In \citet{kangryu15} (Paper I),
we proposed a model for radio {\it gischt} relics in which
a spherical shock sweeps through an elongated cloud of the ICM thermal gas with
an additional population of fossil relativistic electrons.
At the shock, the fossil electrons are re-accelerated to radio-emitting 
energies ($\gamma_e \sim 10^4$) and beyond, producing a diffuse radio source. 
We argued in Paper I that such a model may explain the following characteristics of giant radio relics:
(1) the low occurrence of radio relics compared to the expected frequency of shocks in merging clusters, 
(2) the uniform surface brightness along the length of arc-like relics, 
and (3) the spectral curvature in the integrated radio spectrum that runs over $\sim (0.1-10) \nu_{\rm br}$.
But we were not able to reproduce the abrupt increase of the integrated spectral index, $A_{\nu}$ 
($J_{\nu}\propto \nu^{-A_{\nu}}$), above $\sim$ GHz, 
detected in the observed spectra of some relics including the Sausage relic
\citep[e.g.][]{stroe14,stroe16,trasatti15}

In an effort to explain steep curved radio spectra, in this paper, 
we explore the possibility that the shock breaks out of a finite-size
cloud of fossil electrons, leading to the volume-integrated electron spectrum
much steeper than expected from the simple radiative aging alone.
To that end,
we performed time-dependent, DSA simulations of one-dimensional, spherical
shocks with the parameters relevant for the Sausage relic, which sweep through fossil electron clouds 
of $105-263$ kpc in width.
In the fiducial model, M3.0C1, 
the shock has initially $u_{s,i}\approx 2.8\times 10^3 \kms$ ($M_{s,i}\approx 3.0$),
breaks out of the cloud of 131 kpc after the crossing time of $t_{\rm cross}\approx 45$ Myr, and
decelerates to $u_{s}\approx 2.5\times 10^3 \kms$ ($M_s\approx 2.7$) at  $t_{\rm age}\approx 55$ Myr.
As in Paper I,
we assume that the fossil electron population has a power-law spectrum with exponential cutoff.
We also consider various models with different fossil electron populations or shock Mach numbers,
as summarized in Table 1.
We then calculate the radio surface brightness profile, $I_{\nu}(R)$, and the volume-integrated spectrum, 
$J_{\nu}$, 
adopting the downstream volumes with the same geometrical structure assumed in Paper I.

We find that
a shock of $M_s \approx 2.7-3.0$ and $u_s \approx 2.5-2.8\times 10^3 \kms$ (e.g., the M3.0C1 and M3.3C1 models),
which has exited the fossil electron cloud of 131 kpc about 10 Myr ago, can leave radio-emitting electrons behind,
which produce both 
$I_{\nu}(R)$ and $J_{\nu}$ consistent with the observations reported by \citet{vanweeren10} and \citet{stroe16}.
Although the detailed shape of $J_{\nu}$ depends on the spectrum of fossil electrons (e.g., the slope, $s$,
and the cutoff energy, $\gamma_{e,c}$) as well as the shock Mach number and the magnetic field strength,
the ensuing radio spectrum may explain the steepening of $J_{\nu}$
above $\sim 2$ GHz, seen in the Sausage relic, by adjusting the time since the break-out, $t_{\rm age}-t_{\rm cross}$.

As emphasized in Paper I, the single power-law radio spectrum is 
valid only for a steady planar shock 
with age much longer than 100 Myr (see equation (\ref{fbr})).
For a spherically expanding shock at younger age, the integrated radio spectrum should be curved 
in the range of $\sim (0.1-10)~\nu_{\rm br}$. 
In this study, we further demonstrate that the spectral index could be much bigger than 
$A_{\nu}=\alpha_{\rm shock} + 0.5$ at high frequencies, 
if the shock sweeps out of the fossil electron cloud of finite size.

We conjecture that the typical value of $t_{\rm age}-t_{\rm cross}$ for observed giant radio relics
would be of order of 10 Myr, 
because, if much longer than that, the radio flux density decreases quickly due to fast cooling 
after the shock breaks out of the cloud (see Figures 3 and 4).
In such cases, the spatial offset between the projected shock front and the edge of
the radio relic is of order of 10 kpc, which is
too small to be resolved with currently available observation facilities.
In addition, with the offset, the shock Mach number, $M_{\rm radio}$, derived from the local spectral index, 
$\alpha_{\nu}$, observed at the relic edge
is expected to be {\it slightly} lower than the actual shock Mach number, 
for instance, the Mach number, $M_X$, if the shock can be detected in X-ray observations.  
This is contrary to the observed trend that in some radio relics 
$M_{\rm X} \la M_{\rm radio}$ \citep[e.g.][]{akamatsu13}.
Such observations of $M_{\rm X} \la M_{\rm radio}$, hence,
should be understood by other reasons, for instance, the projection effect of
multiple shock surfaces along line of sights in X-ray and radio observations \citep[e.g.][]{hong15}.

\acknowledgements

We thank the referee for constructive suggestions.
We also thank R. J. van Weeren for comments on the manuscript.
HK was supported by Basic Science Research Program through the National Research Foundation of Korea (NRF) funded by the Ministry of Education (2014R1A1A2057940).
DR was supported by the National Research Foundation of Korea through grant NRF-2014M1A7A1A03029872.

\clearpage

\begin{deluxetable} {lccccccc}
\tablecaption{Models and Simulation Parameters$^a$}
\tablewidth{0pt}
\tablehead{
\colhead {Model} & \colhead{$[M_{s,i}]^b$}  & \colhead{$L_{\rm cloud}$}& \colhead{$[\gamma_{e,c}]^c$} & \colhead{$[s]^c$} 
& \colhead{$[u_{s,i}]^b$}\\
\colhead {Name} & \colhead{ }  & \colhead{(kpc)} & \colhead{ } & \colhead{ } 
& \colhead{$(10^3 \kms)$}
}
\startdata
{\bf M3.0C1} & $3.0$ & 131 & $10^4$ & $4.2$ & $2.8$   \\
{\bf M3.0C1g} & $3.0$ & 131 & $\underline{10^3}$ & $4.2$ & $2.8$   \\
{\bf M3.0C1s} & $3.0$ & 131 & $10^4$ & $\underline{4.0}$ & $2.8$   \\
{\bf M2.5C1} & $\underline{2.5}$ & 131 & $10^4$ &  $4.2$ & $\underline{2.3}$   \\
{\bf M3.3C1} & $\underline{3.3}$ & 131  & $10^4$&  $4.2$ & $\underline{3.1}$    \\
{\bf M4.5C1} & $\underline{4.5}$ & 131 &  $10^4$ & $4.2$ & $\underline{4.2}$   \\ 
{\bf M3.0C2} & $3.0$ & \underline{105} &  $10^4$ & $4.2$ & $2.8$   \\
{\bf M3.0C3} & $3.0$ & \underline{155} &  $10^4$ & $4.2$ & $2.8$  \\
{\bf M3.0C4} & $3.0$ & \underline{263} &  $10^4$ & $4.2$ & $2.8$  \\
{\bf SC1pex1$^d$} & $\underline{2.4}$ & \underline{263} &  $10^4$ & $4.2$ & $\underline{2.2}$  \\
\enddata
\tablenotetext{a}{{\bf M3.0C1} is the fiducial model. The parameters different from those of the fiducial model are marked with underlines. In all the models, $B_1=2.5 \muG$ and $kT_1=3.35$ keV are used.} 
\tablenotetext{c}{Initial shock Mach number and speed.}
\tablenotetext{c}{Fossil electron spectrum: $f_{\rm fossil}\propto p^{-s} \cdot \exp[-(\gamma_e/\gamma_{e,c})^2]$. }
\tablenotetext{d}{Comparison model considered in \citet{kangryu15}.}
\end{deluxetable}

\clearpage

\begin{figure}
\vspace{0cm}
\hspace{-1.3cm}
\includegraphics[scale=0.56]{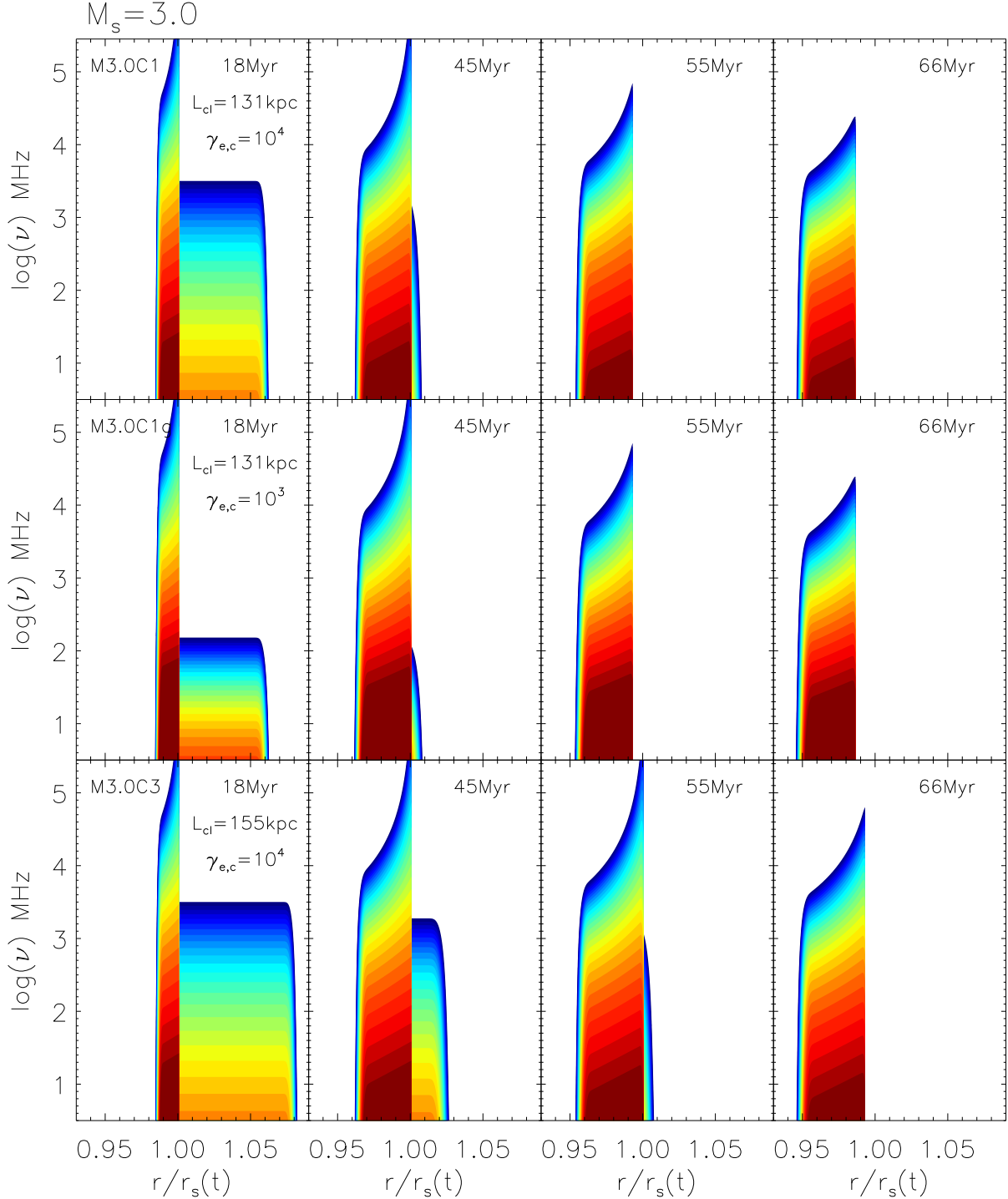}
\vspace{-1cm}
\caption{{Synchrotron emission, $j_{\nu} (r, \nu)$, as a function of radius and frequency,
at four different $t_{\rm age}$ in three models, M3.0C1, M3.0C1g, and M3.0C3 (from top to bottom).
Here, $r/r_s = 1$ marks the shock position.
Red indicates high values and blue indicates low values.
$j_{\nu}$ is displayed in a logarithmic scale, and the value at dark red is about $10^4$ times larger than that at blue.}}
\label{Fig1}
\end{figure}

\clearpage

\begin{figure}
\vspace{-0.7cm}
\hspace{-0.7cm}
\includegraphics[scale=0.87]{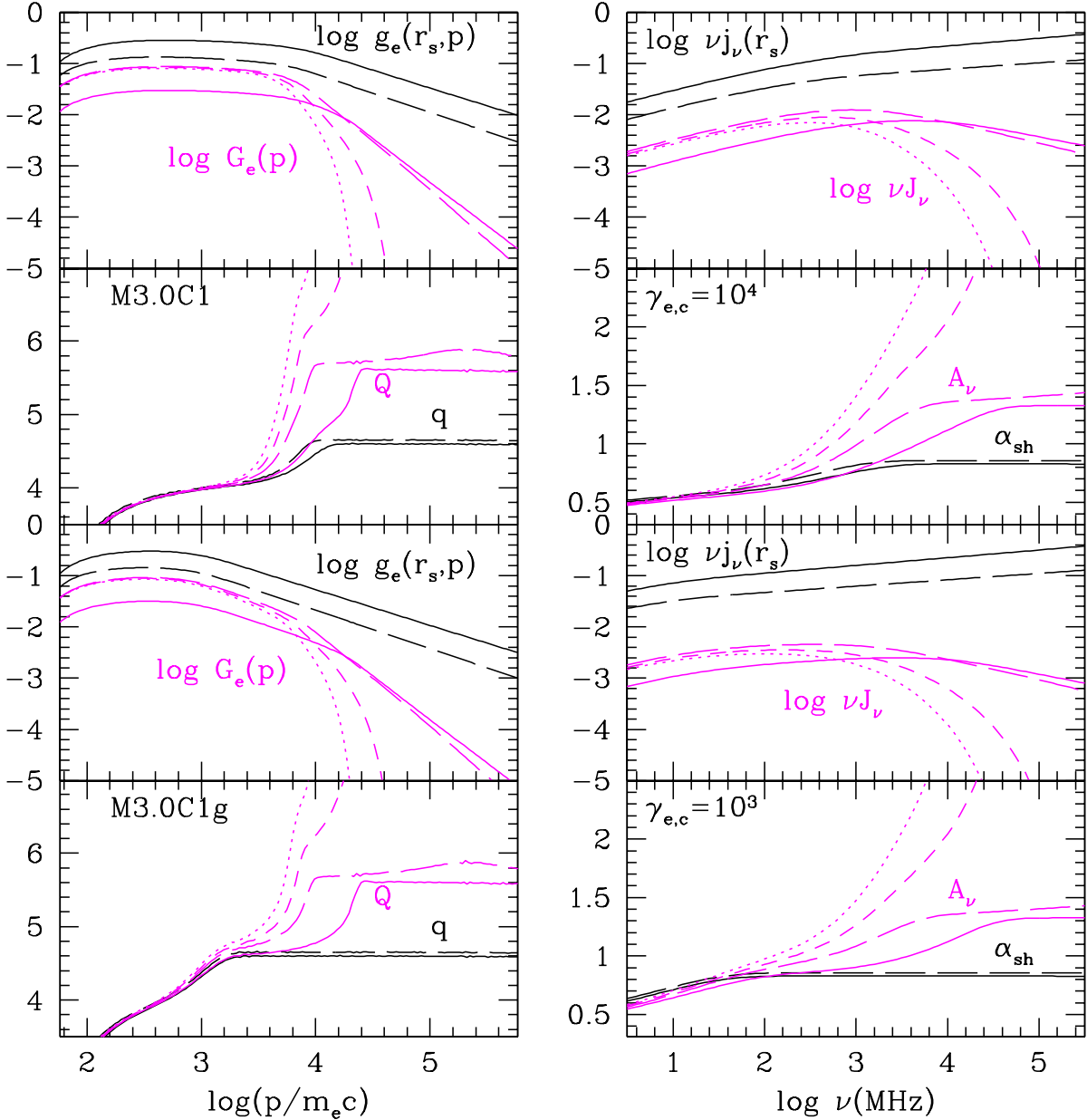}
\vspace{-1cm}
\caption{Left: Electron distribution function at the shock position, $g_e(r_s,p)$ (black lines),
volume-integrated electron distribution function, $G_e(p)$ (magenta lines), and
slopes of electron distribution functions, $q=-d \ln f_e(r_s)/d \ln p$ (black lines) and
$Q=-d \ln F_e/d \ln p$ (magenta lines).
Right: Synchrotron spectrum at the shock position, $\nu j_{\nu}(r_s)$ (black lines),
volume-integrated synchrotron spectrum, $\nu J_{\nu}$ (magenta lines), and
synchrotron spectral indices, $\alpha_{\rm sh}= -d \ln j_{\nu}(r_s)/d \ln \nu $ (black lines)
and $A_{\nu}= -d \ln J_{\nu}/d \ln \nu $ (magenta lines).
The fiducial model M3.0C1 with $\gamma_{e,c}=10^4$ is shown in the upper four panels,
while the model M3.0C1g with $\gamma_{e,c}=10^3$ is shown in the lower four panels. 
Results are shown at $t_{\rm age}=$ 18 (solid lines), 45 (long-dashed), 55 (dashed), and 65 (dotted) Myr.}
\label{Fig2}
\end{figure}

\clearpage

\begin{figure}
\vspace{-0.8cm}
\hspace{-0.5cm}
\includegraphics[scale=0.87]{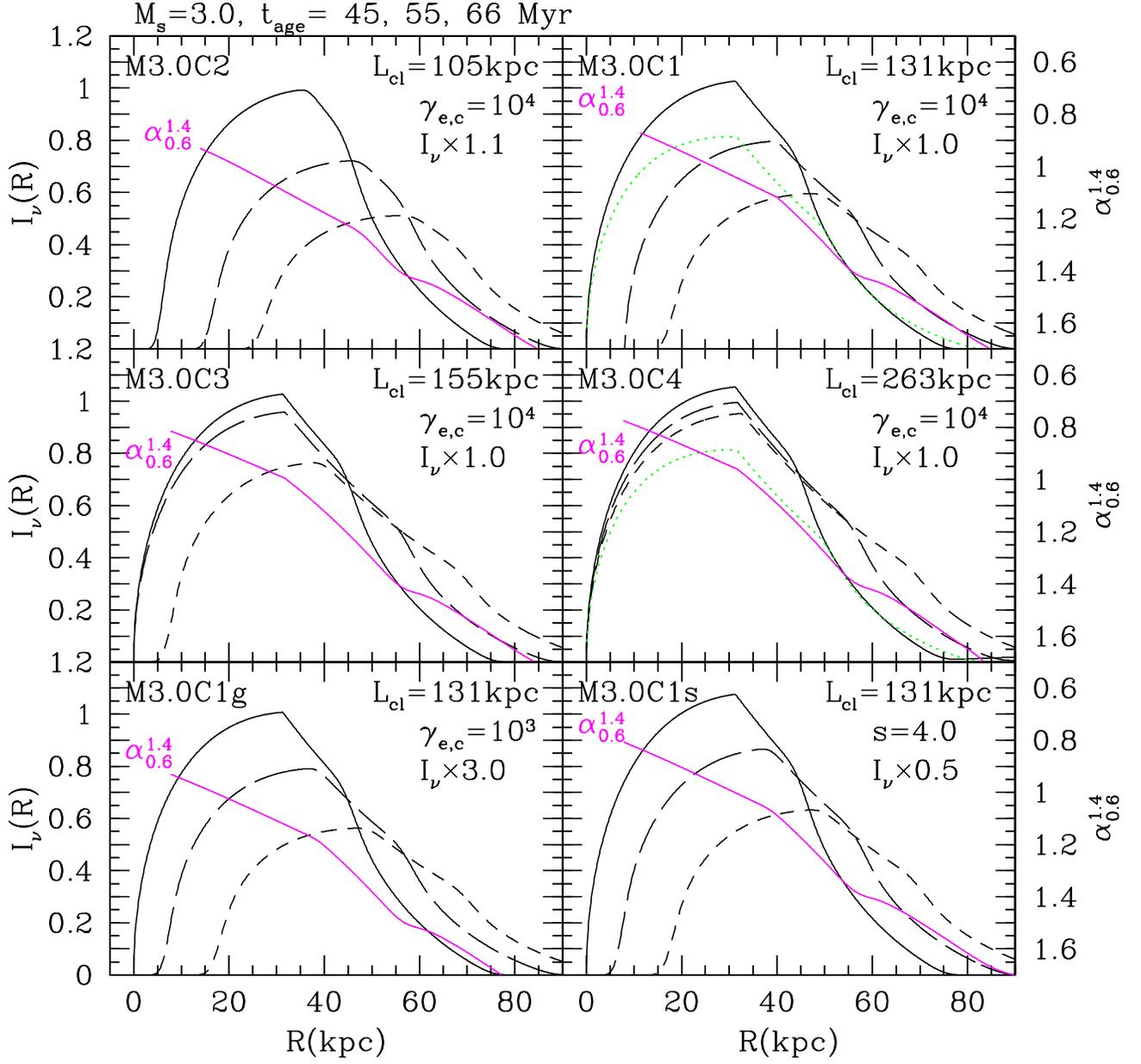}
\vspace{-1cm}
\caption{Surface brightness profile, $I_{\nu}$, at $\nu=0.6$ GHz at
$t_{\rm age}=$ 45 (black solid lines), 55 (long dashed), and 66 Myr (dashed)
in M3.0C2, M3.0C1, M3.0C3, M3.0C4, M3.0C1g, and M3.0C1s models.
That of the comparison model SC1pex1 of Paper I at 62 Myr is also shown (green dotted) in two panels.
See Table 1 for the model parameters.
The radio spectral index, $\alpha_{0.6}^{1.4}$, estimated between 0.6 GHz and 1.4 GHz,
at $t_{\rm age}=55$ Myr (magenta solid lines) is shown as well.
For the extension angle, $\psi = 12^{\circ}$ is adopted.
$I_{\nu}\times X$ is plotted in order to be fitted with one linear scale, where $X$ is specified in each panel.}
\label{Fig3}
\end{figure}

\begin{figure}
\vspace{-1cm}
\hspace{-0.5cm}
\includegraphics[scale=0.87]{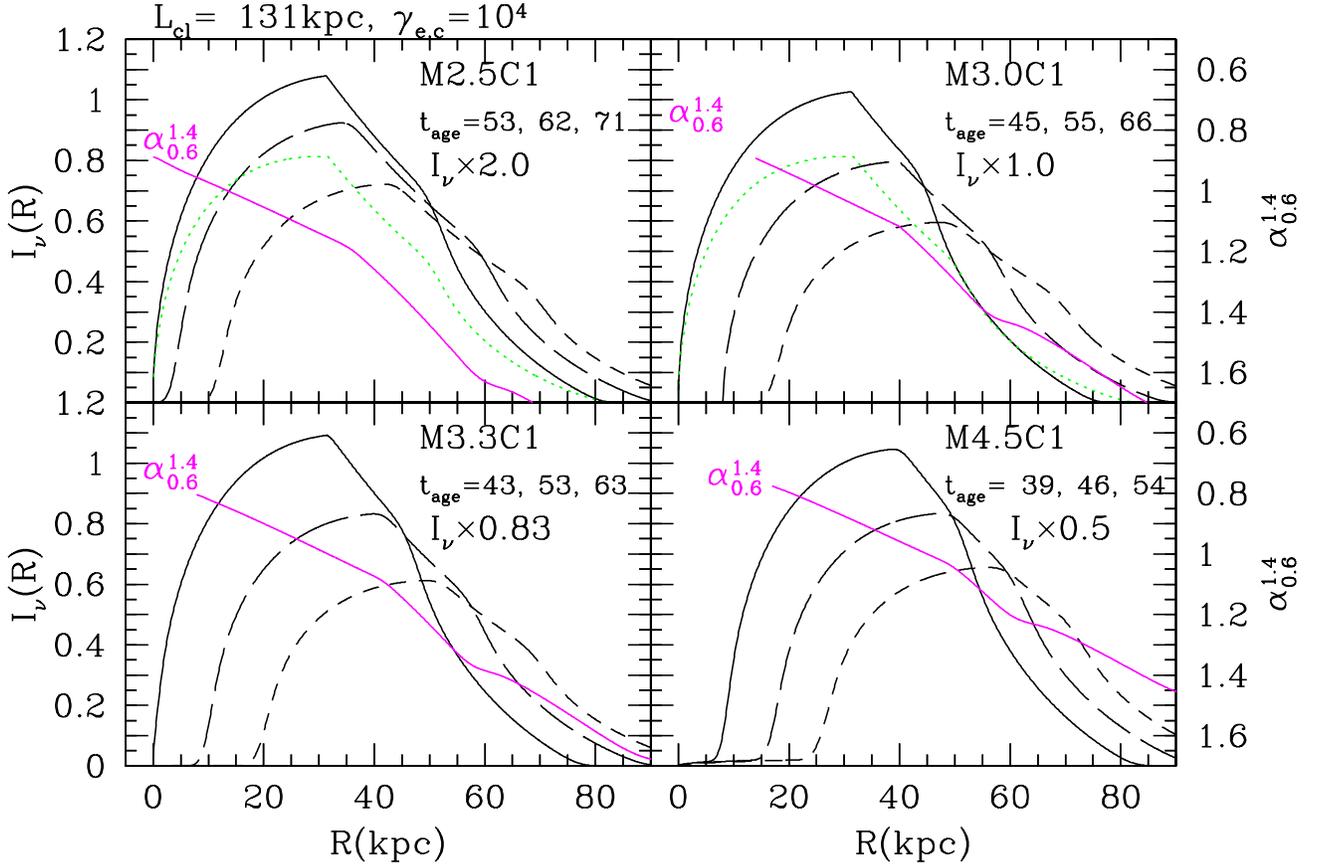}
\vspace{-5.5cm}
\caption{Same as Figure 3 except for the models with different Mach numbers at 
three different $t_{\rm age}$ specified in each panel.
The black solid lines correspond to the earliest ages, while the dashed lines correspond to the latest ages.
$I_{\nu}$ of the model SC1pex1 at 62 Myr is also shown in the green dotted lines in the upper two panels.
See Table 1 for the model parameters.
The radio spectral index, $\alpha_{0.6}^{1.4}$, estimated between 0.6 GHz and 1.4 GHz,
at the middle $t_{\rm age}$ (long-dashed) is shown in the magenta solid lines as well.
The upper left panel for the fiducial model M3.0C1 is the same as that of Figure 3.}
\label{Fig4}
\end{figure}

\clearpage

\begin{figure}
\vspace{-0.8cm}
\hspace{-0.3cm}
\includegraphics[scale=0.87]{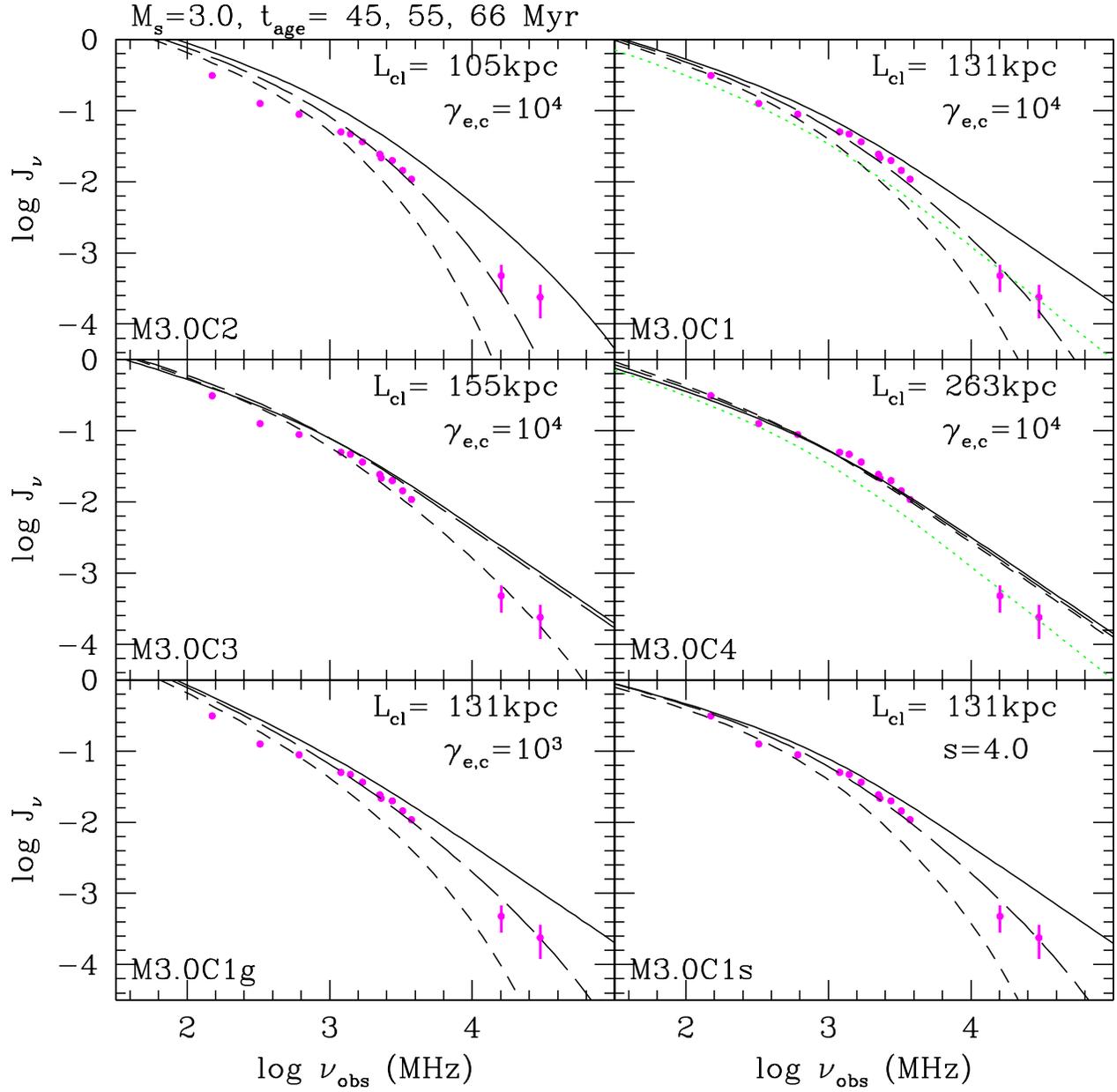}
\vspace{-1cm}
\caption{Volume-integrated synchrotron spectrum, $J_{\nu}$, 
at $t_{\rm age}=$ 45 (solid lines), 55 (long-dashed), and 66 Myr (dashed)
in M3.0C2, M3.0C1, M3.0C3, M3.0C4, M3.0C1g, and M3.0C1s models.
That of the comparison model SC1pex1 of Paper I at 62 Myr is also shown (green dotted) in two panels.
For a qualitative comparison, the observational data in \citet{stroe16}, which are scaled to fit one
of the three curves, are shown with filled magenta circles.
The observational errors are small, about $10 \%$, except for the two
data points at 16 GHz ($42\%$) and 30 GHz ($50\%$).}
\label{Fig5}
\end{figure}

\clearpage

\begin{figure}
\vspace{-1cm}
\hspace{-0.3cm}
\includegraphics[scale=0.87]{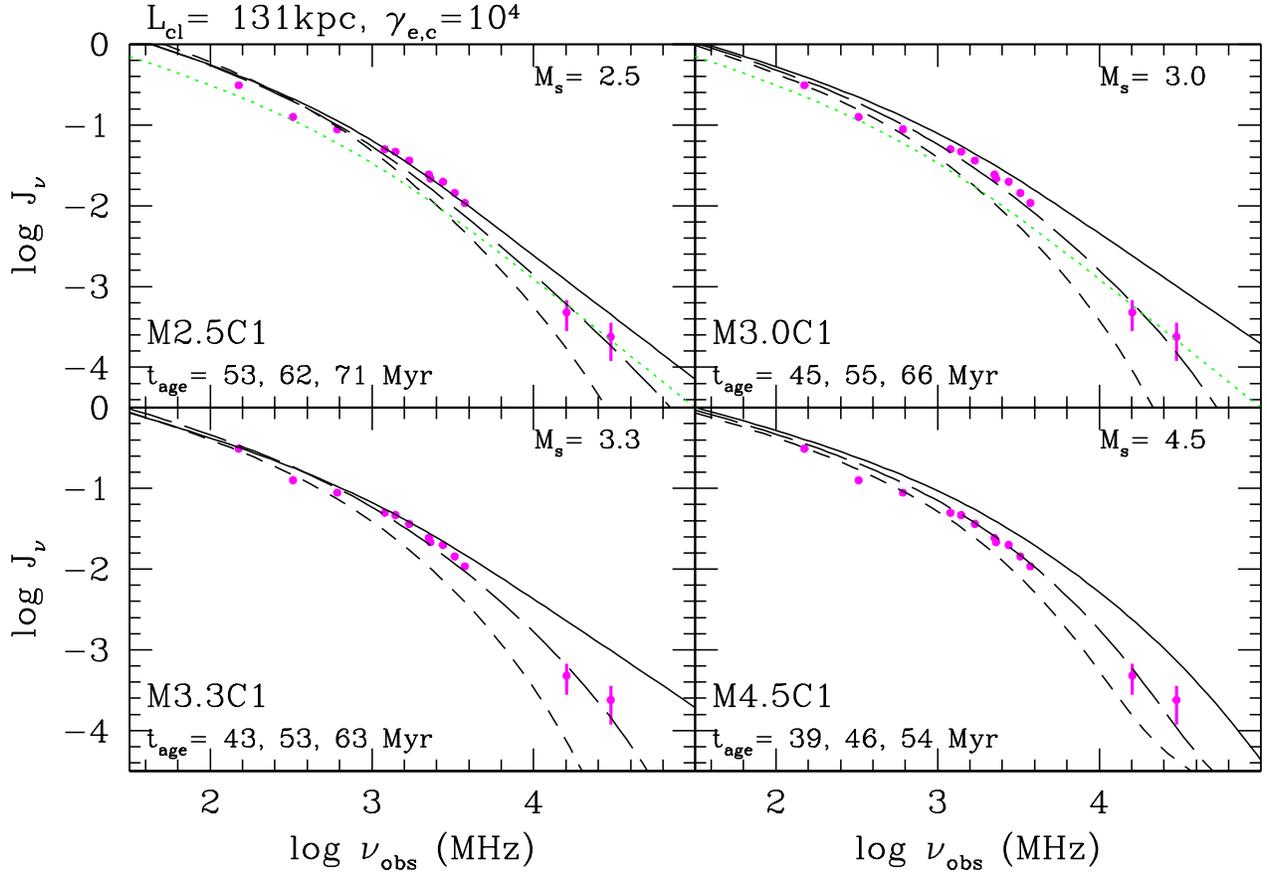}
\vspace{-5.5cm}
\caption{Same as Figure 5 except for the models with different Mach numbers at 
three different $t_{\rm age}$ which are the same as those in Figure 4.
The solid lines correspond to the earliest ages, while the dashed lines correspond to the latest ages.
$J_{\nu}$ of the model SC1pex1 at 62 Myr is also shown in the green dotted lines in the upper two panels.
The upper right panel for the fiducial model M3.0C1 is same as that of Figure 5.}
\label{Fig6}
\end{figure}

\clearpage

\end{document}